\documentclass[twocolumn,showpacs,prb]{revtex4}

\usepackage{graphicx}%
\usepackage{dcolumn}
\usepackage{amsmath}

\begin{document}

\preprint{submitted to Physical Review Letter}
\title{Possible Kondo resonance in PrFe$_4$P$_{12}$ 
studied by bulk-sensitive photoemission}

\author{A. Yamasaki,$^a$ S. Imada,$^a$ T. Nanba,$^b$ A. Sekiyama,$^a$ 
 H. Sugawara,$^c$ H. Sato,$^c$
C. Sekine,$^d$ 
I. Shirotani,$^d$ H. Harima,$^e$ and S. Suga$^a$}
\affiliation{$^a$Graduate School of Engineering Science, Osaka University, 
Osaka 560-8531, Japan\\
$^b$Graduate School of Science and Technology, Kobe University,
Hyogo 657-8501, Japan\\
$^c$Department of Physics, Tokyo Metropolitan University, Tokyo 192-0397, Japan\\
$^d$Department of Electrical and Electronic Engineering,
Muroran Institute of Technology, 
Hokkaido 050-8585, Japan\\
$^e$Institute of Scientific and Industrial Research, Osaka University, Osaka 567-0047, Japan
}

\date{\today}

\begin{abstract}
Pr $4f$ electronic states in Pr-based filled skutterudites ${\rm Pr}T_4X_{12}$
($T$=Fe and Ru; $X$=P and Sb) have been studied by high-resolution bulk-sensitive
Pr $3d\to4f$ resonance photoemission.
A very strong  spectral intensity is observed just below the Fermi level in the heavy-fermion system PrFe$_4$P$_{12}$.  
The increase of its intensity at lower temperatures is observed.
We speculate that this is the Kondo resonance of Pr, the origin of which is attributed to
the strong hybridization between the Pr $4f$ and the conduction electrons.

\end{abstract}


\maketitle
%
%
%
Heavy-fermion properties observed in many Ce and U compounds
and compounds of some other rare-earth elements emerge
when the hybridization between the conduction band in the vicinity of the 
Fermi level and the $f$  state ($c-f$ hybridization) is moderate.
The $4f$ electrons in Pr are more localized and less hybridized with
conduction electrons than in Ce.
No heavy-fermion  Pr compound was known until the discovery of 
${\rm PrInAg_2}$ with a large Sommerfeld
coefficient reaching $\sim$6.5 J/mol K$^2$.~\cite{Yatskar_PrInAg2}
Recently, the heavy electron mass has been found in ${\rm PrFe_4P_{12}}$ under high magnetic 
field.~\cite{Sugawara_dHvA}
In both PrInAg$_2$ and ${\rm PrFe_4P_{12}}$, the crystal-electric field 
ground state is suggested to be a non-Kramers 
doublet,~\cite{Yatskar_PrInAg2,Nakanishi_PFP,Aoki_PFP} which is nonmagnetic 
but has an electric quadrupolar degree of freedom.  
Therefore, the heavy-fermion behaviors in these Pr compounds may result from 
the quadrupolar Kondo effect,~\cite{Cox_Kondo,Kelley_PrInAg2} which was 
first applied to U compounds and is in contrast to the usual spin Kondo effect 
applied to Ce and Yb compounds.  

PrFe$_4$P$_{12}$ is one of the Pr-based filled skutterudites 
${\rm Pr}T_4X_{12}$.  
Among them are ${\rm PrRu_4P_{12}}$ known to show the metal-insulator transition 
at $\sim 64$ K,~\cite{Sekine_MI}  ${\rm PrRu_4Sb_{12}}$ and ${\rm PrOs_4Sb_{12}}$
known as a conventional~\cite{Takeda_Sb} and 
heavy-fermion~\cite{Bauer_POS} superconductor, respectively.
PrFe$_4$P$_{12}$ is particularly interesting due to the phase transition at 
around 6.5 K~\cite{Torikachvili_PFP} and the Kondo-like behaviors.  
Recent studies suggest that the phase transition is associated with the 
ordering of quadrupolar moments.~\cite{Keller,Iwasa_PFP}
In the high-temperature phase, Kondo anomalies are found in the transport 
properties.~\cite{Sato_PFP}
When the low-temperature ordered phase is destroyed by high magnetic field, 
enormously enhanced cyclotron effective mass ($m_{\rm c}^* \simeq 81m_0$) is 
observed in the de Haas-van Alphen measurement.~\cite{Sugawara_dHvA}
A large electronic specific heat coefficient of 
$C_{\rm el}/T\sim 1.2{\rm J/K^2 mol}$ is found under 6 T,~\cite{Sugawara_dHvA}
which suggests the Kondo temperature $T_{\rm K}$ of the order of 10 K.  
These facts suggest the following scenario; 
quadrupolar degree of freedom of the Pr $4f$ state due to the non-Kramers twofold 
degeneracy leads to the quadrupolar Kondo effect, and the 
phase transition 
at 6.5K resulting in the antiquadrupolar ordering 
is driven by the lifting of the quadrupolar degeneracy.  
In order for the quadrupolar Kondo effect to take place, 
$c$--$f$ hybridization must be appreciably strong.  

It has recently been demonstrated that high-resolution photoemission (PE) with use 
of the soft x-ray can reveal bulk electronic states.~\cite{Sekiyama_nature}
The bulk sensitivity owes to the long mean free paths of the high-energy 
photoelectrons.  
Bulk-sensitive measurement must be crucial in the study of Pr $4f$ states 
since the $c$--$f$ hybridization in Ce, Sm, and Yb compounds is known to be much weaker at 
the surface than in the bulk.~\cite{Sekiyama_nature,Iwasaki_CeMX}
In addition, as in the case of Ce systems,
 one needs to enhance the
Pr $4f$ contribution in the PE spectrum by means of resonance photoemission (RPE), 
otherwise the Pr $4f$ state cannot be accurately distinguished from other states.~\cite{Parks_PES,Suga_PES,Kucherenko_PES}

In this paper, we report the results of the bulk-sensitive Pr 3$d\to4f$ RPE measurments
for ${\rm PrFe_4P_{12}}$, ${\rm PrRu_4P_{12}}$, and ${\rm 
PrRu_4Sb_{12}}$.  
It is shown that the Pr $3d\to4f$ RPE spectrum of 
${\rm PrFe_4P_{12}}$ has  much larger spectral weight just 
below the Fermi level ($E_{\rm F}$) than other systems.
We speculate that this spectral weight, 
which increases at lower temperatures, 
comes from the Kondo resonance (KR) due to the $c-f$ hybridization.  
If so, to our knowledge, this is the first observation of the KR in PE of Pr 
systems.

%
Single crystals of ${\rm PrFe_4P_{12}}$ and  ${\rm PrRu_4Sb_{12}}$, and polycrystals of 
${\rm PrRu_4P_{12}}$ were fractured {\it in situ} for 
the soft x-ray absorption (XA) and PE measurements at the BL25SU
of SPring-8.~\cite{Saitoh_BL25SU}
The total energy resolution of the PE measurement was set to $\sim$ 80 meV in 
the high-resolution mode and $\sim$ 130 meV, otherwise.  
The samples were cooled and kept at 20K except for the temperature dependence 
measurement.

%
\begin{figure}
\includegraphics[width=8cm,clip]{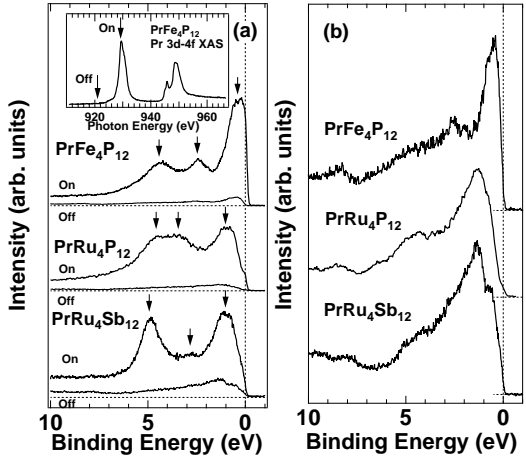}
\caption{
(a) On- and off-RPE spectra normalized by the photon flux.
Inset: Pr $3d\to4f$ XA spectrum for ${\rm PrFe_4P_{12}}$. 
Arrows show the energies at which spectra in the main panel were taken. 
(b) 
Off-RPE spectra taken at 825 eV in an enlarged intensity scale.  
}
\label{Fig_exp}
\end{figure}

The Pr $3d\to4f$ XA spectrum for ${\rm PrFe_4P_{12}}$ is shown in the inset of Fig.\ref{Fig_exp}(a).
This spectrum reflects the predominant ${\rm Pr^{3+}}$ ($4f^2$) character in the initial state.~\cite{Thole_XAS}
Spectra of ${\rm PrRu_4P_{12}}$ and ${\rm PrRu_4Sb_{12}}$ were also quite 
similar to this spectrum.  
Valence-band PE spectra were measured at three photon energies.  
On-RPE spectra were taken at 929.4 eV, around XA maximum.  
Off-RPE spectra were taken at 921 and 825 eV, which were quite similar in 
shape.  
The on- and off- (921.0 eV) RPE spectra are compared in the main panel of 
Fig.\ref{Fig_exp}(a).  
We consider that mainly Pr $4f$ contribution is enhanced in the on-RPE spectra,~\cite{PrOn-Off} and
therefore that the difference between the on- and off-RPE spectra mainly reflects the Pr 
$4f$ spectrum.

\begin{figure}
\includegraphics[width=8cm,clip]{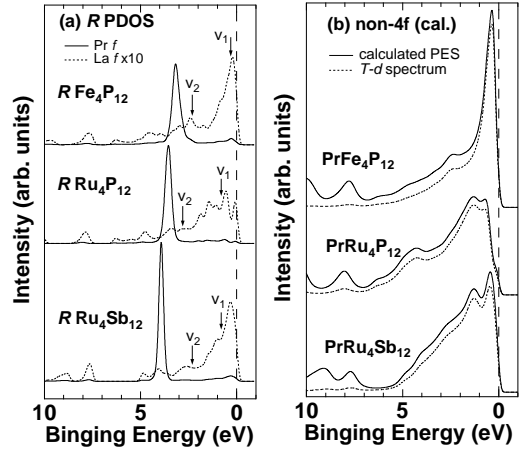}
\caption{
Calculated PE spectra based on band structure calculation.  
Density of states is multiplied by the Fermi-Dirac function for 20~K and 
is broadened by the Gaussian with the full width at half maximum of 80~meV.  
(a)
Calculated Pr (solid line) and La (dashed line: magnified ten times) $f$ 
spectra of Pr$T_4X_{12}$ and La$T_4X_{12}$.  
(b)
Calculated off-RPE spectra (solid lines), where
partial density of states except for Pr $f$ are multiplied by 
the cross sections~\cite{Lindau} and summed up.  
Dashed lines show the contribution of the $T$ $d$ state.  
}
\label{Fig_cal}
\end{figure}

The off-RPE spectra taken at 825 eV with better statistics are shown in 
Fig.\ref{Fig_exp}(b) in a magnified intensity scale.  
The valence band between $E_{\rm F}$ and binding energy ($E_{\rm B}$) of 
$\sim 7$ eV is expected to be composed of 
Pr $5d$ and $4f$,  $T$ $d$, and $X$ $p$ orbitals.  
Among these, main contribution to the off-RPE spectrum (more than 60 \%) 
comes from the $T$ $d$ states according to the photoionization 
cross-section.~\cite{Lindau}
The off-RPE spectral features are reproduced in the theoretical off-RPE spectra based on 
the FLAPW and LDA+U band structure calculation
(see Fig.\ref{Fig_cal}(b)),~\cite{Harima_MI}
where the parameter for the on-site Coulomb interaction $U$ of Pr $4f$ 
electron is set as 0.4 Ry (5.4 eV).  

The on-RPE spectra shown in Fig.\ref{Fig_exp}(a) are characterized by two features.  
First, the on-RPE spectra have various multiple peak structures, where peaks 
(or structures) are indicated by arrows, in 
contrast to the calculated Pr $f$ PDOS (see Fig.\ref{Fig_cal}(a)) that has a strong 
peak and small structures near $E_{\rm F}$ for all the three compounds.  
This feature will be interpreted in the next paragraph taking into account the 
hybridization between the valence band and the Pr $4f$ states 
($v-f$ hybridization) in the 
{\it final} states of PE.  
Second, the intensity near $E_{\rm F}$, {\it i.e.}, between $E_{\rm F}$ and 
$E_{\rm B}\sim 0.3$ eV, is much stronger in PrFe$_4$P$_{12}$ than in other two systems.  
Such strong intensity at $E_{\rm F}$ is neither found in reported Pr $4f$ spectra.  
Later in this paper, this feature will be attributed the strong 
$c-f$ hybridization in the {\it initial} state of PrFe$_4$P$_{12}$.

\begin{figure}
\includegraphics[width=8cm,clip]{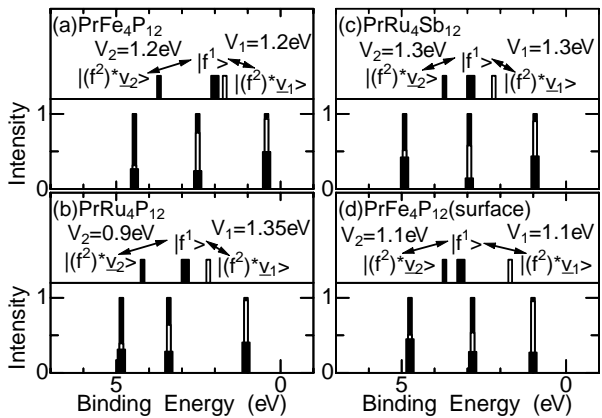}
\caption{
Pr $4f$ spectrum reproduced by the cluster model 
for (a) ${\rm PrFe_4P_{12}}$, 
(b) ${\rm PrRu_4P_{12}}$, (c) ${\rm PrRu_4Sb_{12}}$, and 
(d) surface of ${\rm PrFe_4P_{12}}$.  
Upper panels: Binding energies of bare $|f^1\rangle$ and 
$|(f^2)^*\underline{v_k}\rangle$ final states are shown by the bars and the effective 
hybridization between $|f^1\rangle$ and $|(f^2)^*\underline{v_k}\rangle$ are written.  
Lower panels: The resultant final states are shown.  
The thick filled bars show the weights of the $|f^1\rangle$ state in the final 
states, and correspond to the Pr $4f$ PE line spectrum for the assumed 
$|f^2\rangle$ initial state.  
The thin open and filled bars show the weights of 
$|(f^2)^*\underline{v_1}\rangle$ and $|(f^2)^*\underline{v_2}\rangle$ states, respectively.  
}
\label{Fig_clus}
\end{figure}

Multiple peak structures observed for various Pr compounds have 
been interpreted in terms of the $v-f$ hybridization.~\cite{Parks_PES,Suga_PES,Kucherenko_PES}
We adopt the cluster model,~\cite{Fujimori} {\it i.e.}, the simplified 
version of the single impurity Anderson model (SIAM).~\cite{GS}  
The part of the valence band that hybridizes strongly with the $4f$ state is 
expected to be similar between Pr$T_4X_{12}$ and La$T_4X_{12}$.  
La $f$ PDOS of La$T_4X_{12}$ at a certain energy correspond roughly to the 
$v-f$ hybridization strength at that energy 
since La $f$ states below $E_{\rm F}$ 
comes only from the hybridization with the valence band.  
As a first approximation, we replace the La $f$ PDOS with two levels, 
$v_1$ and $v_2$, the energies of which, $E_{\rm B}(v_k)$, are shown by 
the arrows in Fig.\ref{Fig_cal}(a).  
We now assume that the initial Pr $4f$ state is $|f^2\rangle$.  
Although it turns out later that deviation from this state is appreciable 
in PrFe$_4$P$_{12}$, this is a good approximation when discussing the overall 
spectral features.  
Then the final states of Pr $4f$ PE are linear combinations of 
$|f^1\rangle$, $|(f^2)^*\underline{v_1}\rangle$, and 
$|(f^2)^*\underline{v_2}\rangle$, 
where $\underline{v_k}$ denotes a hole at $v_k$.  
Since the resulting $f^2$ state includes all the excited states, it is denoted 
as $(f^2)^*$ so as to distinguish it from the initial ground state $f^2$.  
The average excitation energy $E((f^2)^*)-E(f^2)$ is $\sim 1.4$ eV 
according to the atomic multiplet calculation.~\cite{Thole_XAS} 
The main origin of this excitation energy is found to be the exchange 
interaction.  
The energies of the bare 
$|(f^2)^*\underline{v_k}\rangle$ with 
respect to the initial state $|f^2\rangle$ are hence 
$E_{\rm B}(v_k)+[E((f^2)^*)-E(f^2)]$ and are shown by the thin open and filled 
bars in the upper pannels of 
Figs.\ref{Fig_clus}~(a)-(c).  
We take the remaining three parameters, $E_{\rm B}$ of the bare 
$|f^1\rangle$ ($E_0$), the hybridization between 
$|f^1\rangle$ and $|(f^2)^*\underline{v_k}\rangle$ ($V_k$), to be free 
parameters, and numerically solve the $3 \times 3$ Hamiltonian matrix.  
When the parameters are set as in the upper pannels of Figs.\ref{Fig_clus} (a)-(c), 
the three final states are obtained as shown in the lower pannels. 
At each of the three eigen-energies of the final states is placed a set of vertical 
bars the lengths of which are proportional to the weights of $|f^1\rangle$ 
(thick filled bar), $|(f^2)^*\underline{v_1}\rangle$ (thin open bar), and 
$|(f^2)^*\underline{v_2}\rangle$ (thin filled bar).  
Since we assume that the initial state is $|f^2\rangle$, the Pr $4f$ 
excitation intensity is proportional to the weight of 
the $|f^1\rangle$ in each final state.  
Therefore, the thick filled bars show the obtained line spectrum.  
The line spectra qualitatively well reproduce the experimentally 
observed system dependence in the energy positions and intensity ratios of the 
three-peak structures of the on-RPE spectra (see Fig.\ref{Fig_exp}(a)).  
The present analysis revealed the character of each final 
states.  
For example, the final state with the smallest $E_{\rm B}$ is the 
bonding state between $|f^1\rangle$ and $|(f^2)^*\underline{v_1}\rangle$.
The trend in the E$_B$ of bare   $|f^1\rangle$ corresponds to some extent with the trend in the
peak position of Pr $f$ PDOS in Fig.\ref{Fig_cal}(a).
The origin of these trends could be that E$_B$ of the Pr $4f$ electron becomes smaller because
the negative $X$ ion comes closer to Pr atom in the direction of PrRu$_4$Sb$_{12}$, 
PrRu$_4$P$_{12}$, PrFe$_4$P$_{12}$.

The present Pr $4f$ spectrum 
of ${\rm PrFe_4P_{12}}$ obtained from the bulk-sensitive $3d\to4f$ RPE 
is qualitatively different from that obtained from the surface-sensitive 
$4d\to4f$ RPE.~\cite{Ishii}
The surface-sensitive spectrum also has a three peak structure but the peak at 
$E_{\rm B}\sim$ 4.5 eV is the strongest and the intensity at $E_{\rm F}$ is 
negligible.  
The origin of the difference is the increase of the localization of $4f$ 
electrons at the surface, in other words, the increase of the $4f$ binding 
energy and the decrease of the hybridization.  
By making such changes in $E_0$ and $V_k$, 
the surface-sensitive spectrum is reproduced (see Fig.\ref{Fig_clus}(d)).  

\begin{figure}
\includegraphics[width=8cm,clip]{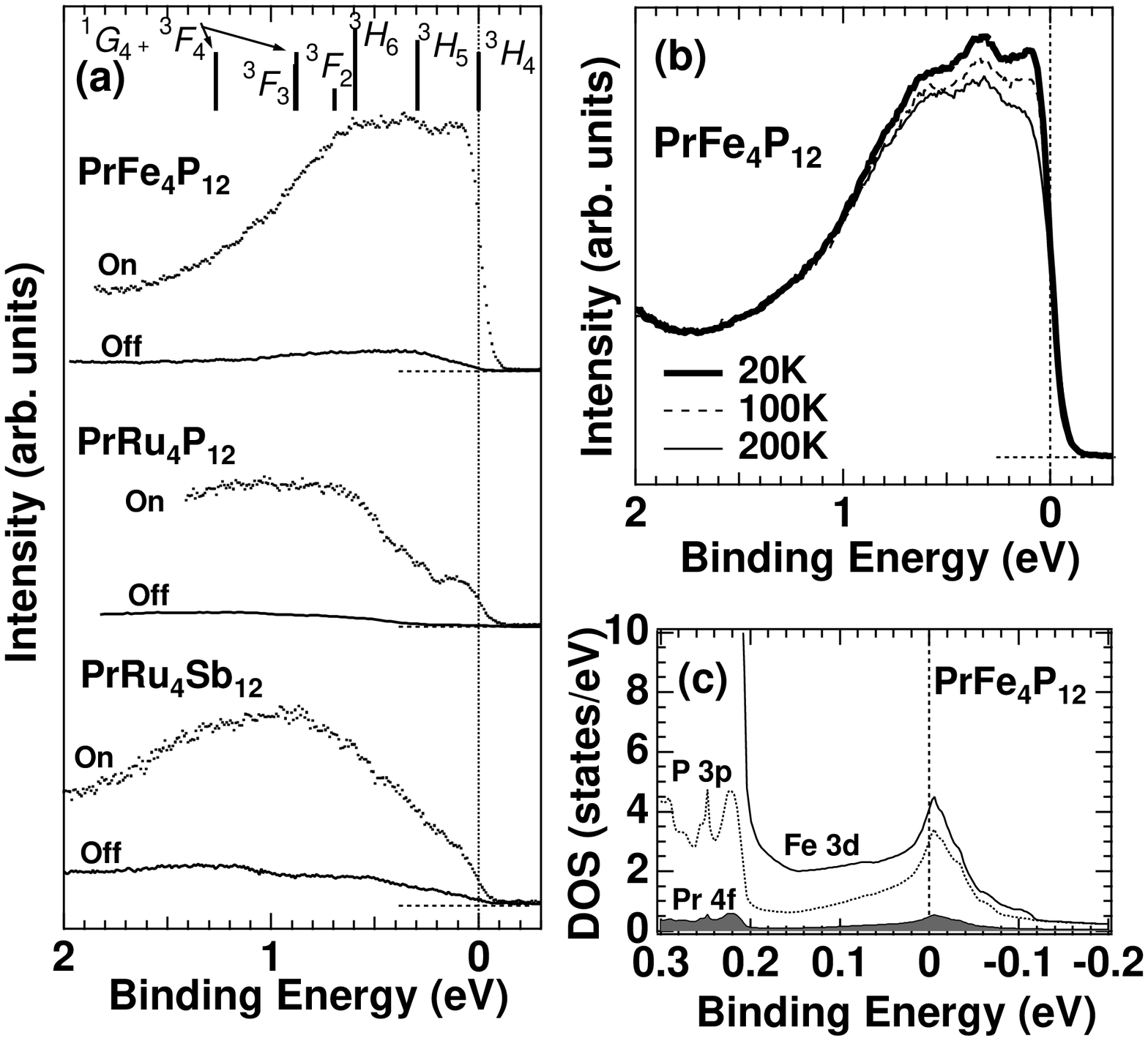}
\caption{
(a)
High-resolution Pr $3d\to4f$ on- (dots) and off- (solid lines) RPE spectra 
near $E_{\rm F}$ at 20~K.
The vertical lines show the energy positions of the atomic $4f^2$ multiplets 
with the ground state set at $E_{\rm F}$.  
(b)
Temperature dependence of the on-RPE spectrum of ${\rm PrFe_4P_{12}}$.  
(c)
Calculated  partial density of states.  
}
\label{Fig_EF}
\end{figure}

We  measured  the Pr $3d\to4f$ RPE 
spectra near $E_{\rm F}$ with high resolution as shown in  Fig.\ref{Fig_EF}(a).
The most prominent feature  is the strong peak of 
${\rm PrFe_4P_{12}}$ at $E_{\rm B}\simeq100$ meV.  
The Pr $4f$ spectra of ${\rm PrRu_4P_{12}}$ and ${\rm PrRu_4Sb_{12}}$, 
on the other hand, 
decrease continuously with some humps as approaching $E_{\rm F}$. 
Spectral features similar to ${\rm PrRu_4P_{12}}$ and ${\rm PrRu_4Sb_{12}}$ 
has been found for very localized Ce systems such as CePdAs, 
in which Ce $4f$ 
takes nearly pure $4f^1$ state.~\cite{Iwasaki_CeMX}
This indicates that pure $4f^2$ state is realized in 
${\rm PrRu_4P_{12}}$ and ${\rm PrRu_4Sb_{12}}$.  
On the other hand, similarity between the ${\rm PrFe_4P_{12}}$'s and Kondo Ce 
compound's spectra \cite{Sekiyama_nature} suggests that the Pr $4f^2$-dominant Kondo state, 
with the finite contribution of $4f^1$ or $4f^3$ state, 
is formed in ${\rm PrFe_4P_{12}}$.  

The present energy resolution of $\sim 80$ meV exceeds the characteristic 
energy $k_{\rm B}T_{\rm K} \sim 1$ meV for ${\rm PrFe_4P_{12}}$.  
KR has been observed even in such cases, 
for example, for ${\rm CeRu_2Si_2}$ 
($T_{\rm K}\sim 20$ K) \cite{Sekiyama_nature} and YbInCu$_4$ 
($T_{\rm K} \sim 25$ K for $T > 42$ K) \cite{Sato_Yb} 
with energy resolution of $\sim 100$ meV.  

In the Kondo Ce (Yb) system, the KR is accompanied by the 
spin-orbit partner, the $E_{\rm B}$ of which corresponds to the spin-orbit 
excitation, $J=5/2\to 7/2$ ($J=7/2\to 5/2$), of the $4f^1$ 
($4f^{13}$)-dominant state.~\cite{GS}  
A KR in Pr would then be accompanied by satellites corresponding 
to the excitation from the ground state ($^3H_4$) to excited states 
($^3H_5$, $^3H_6$, $^3F_2$, and so on) of the $4f^2$ states.  
Fig.\ref{Fig_EF}(a) shows that the on-RPE spectrum of ${\rm PrFe_4P_{12}}$ have 
strucures at $\sim 0.3$ and $\sim$0.6 eV which correspond to the lowest few 
excitation energies.

KR is expected to depend upon temperature reflecting the 
temperature dependence of the $4f$ occupation number.  
In fact, a temperature dependence was found as the temperature 
approaches the suggested $T_{\rm K}\sim 10$ K as shown 
in Fig.\ref{Fig_EF}(b).  
The temperature dependence was reproducible in both heat-up and cool-down
processes.  
The temperature dependence is characterized not only by the narrowing of the $\sim$ 0.1 eV structure but also by the increase of the weight of all the 
structures at $\sim$ 0.1, $\sim$ 0.3, and $\sim$ 0.6 eV.  Although the former can be 
attributed at least partly to the thermal broadening, the latter should be 
attributed to intrisic temperature dependence of the excitation spectrum.  
Therefore it is quite possible that the $\sim$ 0.1 eV structure is the KR and 
the $\sim$ 0.3 and $\sim$ 0.6 structures are its satellite structures.  

The temperature dependence can be a vital clue to check 
whether the observed structure is the Kondo peak itself or the tail of the 
Kondo peak centered above $E_{\rm F}$.  
These cases correspond respectively to the $c_2|f^2\rangle + c_3|f^3\rangle$ 
or $d_1|f^1\rangle + d_2|f^2\rangle$ initial states, where the hole or 
electron in the valence or conduction band is not denoted explicitly.  
The non-crossing approximation (NCA) calculation based on the SIAM for the Ce system \cite{Kasai_NCA} shows that, as temperature is lowered, the Kondo tail 
is sharpened \cite{Reinert_CeCu2Si2} but the {\it weights} of both the Kondo 
tail and its spin-orbit partner {\it decrease} 
when the spectra are normalized in a similar way as in Fig.\ref{Fig_EF}(b).  
This contradicts with the present temperature dependence for 
${\rm PrFe_4P_{12}}$.  
On the other hand, for Yb systems, it is well known that the intensities of 
both the Kondo 
peak itself and its spin-orbit partner increase with decreasing temperature.~\cite{Tjeng_Yb}  
Since this is consistent with the ${\rm PrFe_4P_{12}}$'s temperature 
dependence,  we tend to believe that the observed structure is the Kondo peak 
itself, and therefore that the initial state is dominated by 
$c_2|f^2\rangle + c_3|f^3\rangle$.  
We consider that the Kondo peak at around $k_{\rm B}T_{\rm K}\sim 1$ meV 
is broadened due to the energy resolution of $\sim 80 $ meV resulting in the 
observed structure at $\sim 100$ meV.

Microscopic origin of the $c-f$ hybridization 
is considered to be the P $3p-$Pr $4f$ mixing since the nearest 
neighbors of the Pr atom are the twelve P atoms. 
The large coordination number definitely enhances the effective $p-f$ mixing.  
It has been pointed out that the calculated P $p$ PDOS of 
$R$Fe$_4$P$_{12}$ shows a sharp peak in the vicinity of 
$E_{\rm F}$.~\cite{Sugawara_FS,Harima_MI} 
This is also the case for PrFe$_4$P$_{12}$ as shown in Fig.\ref{Fig_EF}(c). 
Therefore, the large P $3p$ PDOS at $E_{\rm F}$ together with the large effective 
P $3p-$Pr $4f$ mixing is interpreted to be the origin of the Kondo state in 
${\rm PrFe_4P_{12}}$.

In conclusion, our analysis of the data suggests that there may be a Kondo resonance (KR) in the $4f$
photoemission spectrum of PrFe$_4$P$_{12}$, whereas no KR was 
seen in 
PrRu$_4$P$_{12}$ and PrRu$_4$Sb$_{12}$.  
The origin of the KR in PrFe$_4$P$_{12}$ is considered to be 
the Kondo effect caused by  the strong hybridization 
between the Pr $4f$ and P $3p$ states in the vicinity of $E_{\rm F}$.

We would like to thank Profs.\ O. Sakai and K. Miyake for fruitful discussions.  
The research was performed at SPring-8  (Proposal Nos.~2001A0158-NS-np and 2002A0433-NS1-np) 
under the support of
a Giant-in-Aid for COE Research (10CE2004) and Scientific 
Research Priority Area "Skutterudite" (No.15072206)  of the Ministry of
Education, Culture, Sports, Science, and Technology, Japan.


\begin{references}


\bibitem{Yatskar_PrInAg2}
A. Yatskar, W. P. Beyermann, R. Movshovich, and P. C. Canfield,
Phys. Rev. Lett. {\bf 77}, 3637 (1996).

\bibitem{Sugawara_dHvA}
H. Sugawara, T. D. Matsuda, K. Abe, Y. Aoki, H. Sato, S. Nojiri, Y. Inada, R. Settai, and Y. \={O}nuki,
Phys. Rev. B {\bf 66}, 134411 (2002).

\bibitem{Nakanishi_PFP}
Y. Nakanishi, T. Simizu, M. Yoshizawa, T. D. Matsuda, H. Sugawara, and H. Sato,
Phys. Rev. B {\bf 63}, 184429 (2001).

\bibitem{Aoki_PFP}
Y. Aoki, T. Namiki, T. D. Matsuda, K. Abe, H. Sugawara, and H. Sato,
Phys. Rev. B {\bf 65}, 064446 (2002).

\bibitem{Cox_Kondo}
D. L. Cox,
Phys. Rev. Lett. {\bf 59}, 1240 (1987).

\bibitem{Kelley_PrInAg2}
T. M. Kelley, W. P. Beyermann, R. A. Robinson, F. Trouw, P. C. Canfield, and H. Nakotte,
Phys. Rev. B {\bf 61}, 1831 (2000).

\bibitem{Sekine_MI}
C. Sekine, T. Uchiumi, I. Shirotani, and T. Yagi,
Phys. Rev. Lett. {\bf 79}, 3218 (1997).

\bibitem{Takeda_Sb}
N. Takeda and M. Ishikawa,
J. Phys. Soc. Jpn. {\bf 69}, 868 (2000).
\bibitem{Bauer_POS}
E. D. Bauer, N. A. Frederick, P.-C. Ho, V. S. Zapf, and M. B. Maple, 
Phys.\ Rev.\ B {\bf 65}, 100506R (2002).

\bibitem{Torikachvili_PFP}
M. S. Torikachvili, J. W. Chen, Y. Dalichaouch, R. P. Guertin, M. W. McElfresh,
C. Rossel, M. B. Maple, and G. P. Meisner,
Phys. Rev. B {\bf 36}, 8660 (1987).

\bibitem{Iwasa_PFP}
K. Iwasa, Y. Watanabe, K. Kuwahara, M. Kohgi, H. Sugawara, T. D. Matsuda, Y. Aoki, and H. Sato,
Physica B  {\bf 312-313}, 834 (2002).

\bibitem{Keller}
L. Keller, P. Fischer, T. Herrmannsdorfer, A. Donni, H. Sugawara,
T. D. Matsuda, K. Abe, Y. Aoki, and H. Sato,
J. Alloys Compd. {\bf 323-324}, 516 (2001).

\bibitem{Sato_PFP}
H. Sato, Y. Abe, H. Okada, T. D. Matsuda, K. Abe, H. Sugawara, and Y. Aoki, 
Phys. Rev. B {\bf 62}, 15125 (2000).

\bibitem{Sekiyama_nature}
A. Sekiyama, T. Iwasaki, K. Matsuda, Y. Saitoh, Y. \={O}nuki, and S. Suga,
Nature (London) {\bf 403}, 396 (2000).


\bibitem{Iwasaki_CeMX}
T. Iwasaki, A. Sekiyama, A. Yamasaki, M. Okazaki, K. Kadono, H. Utsunomiya, S. Imada, Y. Saitoh, T. Muro, T. Matsushita,
H. Harima, S. Yoshii, M. Kasuya, A. Ochiai, T. Oguchi, K. Katoh, Y. Niide, K. Takegahara, and S. Suga,
Phys. Rev. B {\bf 65}, 195109 (2002).

\bibitem{Parks_PES}
R. D. Parks, S. Raaen, M. L. denBoer, Y.-S. Chang, and G. P. Williams,
Phys. Rev. Lett. {\bf 52}, 2176 (1984).

\bibitem{Suga_PES}
S. Suga, S. Imada, H. Yamada, Y. Saitoh, T. Nanba, and S. Kunii,
Phys. Rev. B {\bf 52}, 1584 (1995).

\bibitem{Kucherenko_PES}
Yu. Kucherenko, M. Finken, S. L. Molodtsov, M. Heber, J. Boysen, C. Laubschat, and G. Behr,
Phys. Rev. B {\bf 65}, 165119 (2002).

\bibitem{Saitoh_BL25SU}
Y. Saitoh, H. Kimura, Y. Suzuki, T. Nakatani, T. Matsushita, T. Muro, T. Miyahara, 
M. Fujisawa, S. Ueda, H. Harada, M. Kotsugi, A. Sekiyama, and S. Suga,
Rev. Sci. Instrum. {\bf 71}, 3254 (2000).

\bibitem{Thole_XAS}
B. T. Thole, G. van der Laan, J. C. Fuggle, G. A. Sawatzky, R. C. Karnatak,
and J.-M. Esteva,
Phys. Rev. B {\bf 32}, 5107 (1985).

\bibitem{PrOn-Off}  
In the on-RPE spectra, not only Pr $4f$ compounent but also other components 
can be enhanced.~\cite{Olson_La,Molodtsov_La}
The latter may include Pr $5d$, $T$ $d$, and $X$ $p$ states.  
Contribution from the Pr $5d$ state can be estimated from the La $3d\to 4f$ RPE 
of La$T_4X_{12}$ (not shown) to be at most of the order of the off-RPE 
intensity of PrFe$_4$P$_{12}$ (see Fig.\ref{Fig_exp}(a)).  
Also the cross sections of the non-Pr $4f$ components change between on- and 
off-RPE conditions but these changes are at most 5 \% of the off-RPE intensity.
There might be further enhancement of the non-Pr $4f$ state due to the 
interplay of the $c-f$ mixing and the cross term between the Pr $4f$ and other 
excitation processes.  
However, since the matrix element of Pr $4f$ is far larger than those of other 
states judging from the large enhancement, the cross term should be much 
smaller than the pure Pr $4f$ term.  

\bibitem{Olson_La}
C. G. Olson, P. J. Benning, M. Schmidt, D. W. Lynch, P. Canfield, and D. M. Wieliczka,
Phys. Rev. Lett. {\bf 76}, 4265 (1996).

\bibitem{Molodtsov_La}
S. L. Molodtsov, M. Richter, S. Danzenb\"acher, S. Wieling, L. Steinbeck, and C. Laubschat,
Phys. Rev. Lett. {\bf 78}, 142 (1997).

\bibitem{Lindau}
J. J. Yeh and I. Lindau,
Atomic Data and Nuclear Data Tables {\bf 32}, 1 (1985).


\bibitem{Harima_MI}
H. Harima and K. Takegahara,
Physica B  {\bf 312-313}, 843 (2002).

\bibitem{Fujimori}
A. Fujimori, 
Phys. Rev. B {\bf 27}, 3992 (1983). 

\bibitem{GS}
O. Gunnarsson and K. Sch\"onhammer, 
Phys. Rev. B {\bf 28}, 4315 (1983).

\bibitem{Ishii}
H. Ishii, K. Obu, M. Shinoda, C. Lee, and Y. Takayama,
J. Phys. Soc. Jpn. {\bf 71}, 156 (2002).

\bibitem{Sato_Yb}
H. Sato, K. Hiraoka, M. Taniguchi, Y. Takeda, M. Arita, K. Shimada, H. Namatame, 
A. Kimura, K. Kojima, T. Muro, Y. Saitoh, A. Sekiyama, and S. Suga,
J. Synchrotron Rad. {\bf 9}, 229 (2002).

\bibitem{Kasai_NCA}
S. Kasai, S. Imada, A. Sekiyama, and S. Suga,
unpublished.  

\bibitem{Reinert_CeCu2Si2}
F. Reinert, D. Ehm, S. Schmidt, G. Nicolay, S. H\"ufner, J. Kroha, O. Trovarelli, and C. Geibel,
Phys.\ Rev.\ Lett.\ {\bf 87}, 106401 (2001).

\bibitem{Tjeng_Yb}
L. H. Tjeng, S.-J. Oh, E.-J. Cho, H.-J. Lin, C. T. Chen, G.-H. Gweon, J.-H. Park, J. W. Allen, and T. Suzuki, 
Phys. Rev. Lett. {\bf 71}, 1419 (1993).

\bibitem{Sugawara_FS}
H. Sugawara, Y. Abe, Y. Aoki, H. Sato, M. Hedo, R. Settai, Y. \={O}nuki, and H. Harima,
J. Phys. Soc. Jpn. {\bf 69}, 2938 (2000).

\end{references}
\end{document}